\documentstyle[12pt,psfig]{article}
\begin{document}

\title{Analysis of Sunspot Number Fluctuations}

\author{{\em H. Fanchiotti,
 S. J. Sciutto\/}\footnote{E-mails:
huner@fisica.unlp.edu.ar;
 sciutto@fisica.unlp.edu.ar} \\
Laboratorio de F\'{\i}sica Te\'{o}rica\\
Departamento de F\'{\i}sica and IFLP (CONICET),\\
Universidad Nacional de La Plata \\
C.C. 67 - 1900 La Plata, Argentina.\\*[3pt]
{\em C.A. Garc\'{\i}a Canal\/}\footnote{E-mail:Carlos.A.Garcia@uv.es. Also
at Laboratorio de F\'{\i}sica Te\'{o}rica, Departamento de
F\'{\i}sica and IFLP (CONICET), Universidad Nacional de La Plata.}\\
Departamento de F\'{\i}sica Te\'{o}rica\\Universidad de
Valencia\\E-46100, Burjassot, Valencia, Spain.\\*[3pt]
{\em C. Hojvat\/}\footnote{E-mail: hojvat@fnal.gov}\\ Fermilab\\
 P. O. Box 500 -- Batavia IL 60510-0500,
 USA}
\date{}
\maketitle
\begin{center}
{\bf Abstract}\end{center} { Monthly averages of the sunspot
number visible on the sun, observed from
 1749, Zurich Observatory, and from 1848 other observatories, have been
 analyzed.  This time signal presents a frequency power spectra with a
 clear $1/f^\alpha$ behavior with $\alpha\simeq 0.8 \pm 0.2$. The well
 known cycle of approximately 11 years, clearly present in the spectrum,
 does not produce a sensible distortion of that behavior. The eventual
 characterization of the sunspot time series as a fractal is analyzed
 by means of the detrended fluctuation analysis. The jump size distribution
 of the signal is also studied.}

\section{Introduction}
A sunspot \cite{bray} is a relatively cool area on the solar face
that appears as a dark blemish. The number of sunspots is
continuously changing in time in a random fashion and constitutes a
typical, a priori, random time series. Because of the symmetry of
the twisted magnetic lines that originate sunspots, they are
generally seen in pairs or in groups of pairs at both sides of the
solar equator. As the sunspot cycle progresses, spots appear
closer to the sun's equator giving rise to the so called
``butterfly diagram'' in the latitude distribution. Magnetic
fields about the sunspots are very strong and keep heat out of
these regions on the sun surface. They are formed when magnetic
field lines are twisted and poke through the solar photosphere.
The twisted magnetic fields above sunspots are sites where solar
flares are observed. It has been found that chromospheric flares
show a very close statistical relationship with sunspots. This is
illustrated by the Waldmeir relationship \cite{bray} between the
mean number of flares per day, $E$, and the sunspot number, $R$,
namely $E = 0.061\,R$. Being the number of sunspots much larger
than the corresponding to flares, it seems more appropriate for
statistical analysis. Moreover, this connection between sunspots
and solar magnetic field is particularly motivating for a further
analysis of their frequency of appearance. In summary, being a
consequence of the solar cycle, the variation of the number of
sunspots is closely related to the so-called solar activity. In
fact, the solar maximum activity corresponds to many sunspots
present, while the minimum of activity is related to very few
sunspots.

Another effect of the solar activity is the presence of
fluctuations in the interplanetary magnetic field. Experimental
data suggest \cite{matt} a $1/f$ dependence for the frequency
spectra of its fluctuations in the range of large times, namely,
$2.7\times10^{-6}$ Hz to $1.8\times 10^{-5}$ Hz. In this analysis
it is argued that on the basis of the observed behavior there is a
superposition of signals with scale invariant distributions of
correlation times. In other words, a kinematical superposition of
signals due to scale-invariant reconnection of magnetic
structures, near the solar surface, gives rise to the $1/f$
spectrum. It should be also noticed  that at higher frequencies
the interplanetary spectrum is known to have approximatively an
$1/f^{5/3}$ behavior; this time associated with
magnetohydrodynamics (MHD) homogeneous turbulence.

The result of our analysis of the sunspot number distribution
along almost 250 years shows a clear correlation among all these
phenomena.

After introducing the data and the different methods of analysis
used, we present the results with a further comment on their
implications.

\section{Sunspot Data}

There is experimental evidence and data registers on sunspots
since Galileo (1610). Afterwards, the sunspot data were provided
by the Z\"urich Observatory from 1749 to 1981. Since January 1981,
the data are provided by the Sunspot Index Data Center (SIDC)
\cite{data}.

\begin{figure}[htbp]
\vspace*{13pt}
\centerline{\psfig{file=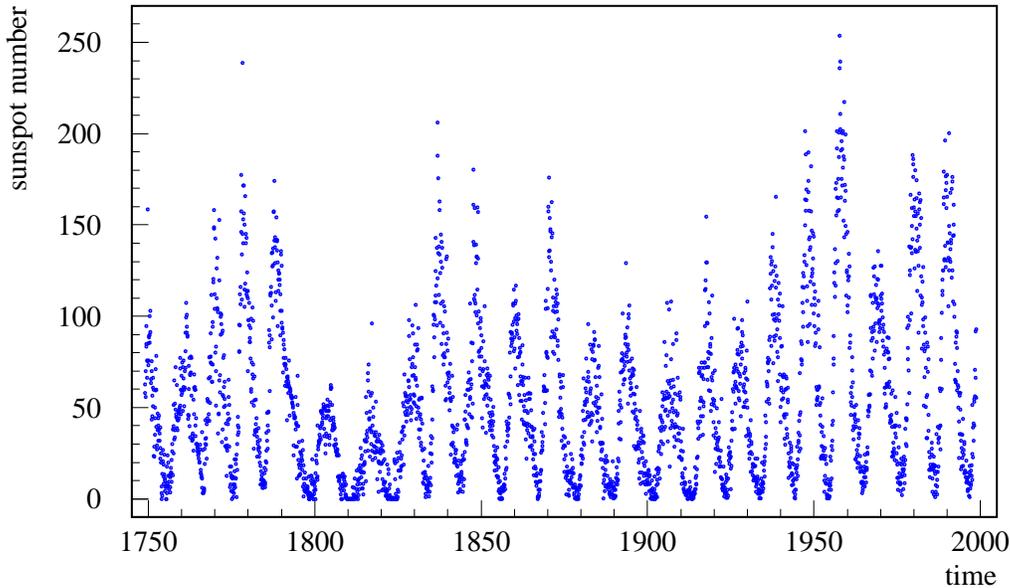}}
\caption{Observed spots number as a function of time}
\label{FIG:nsvst}
\end{figure}

In Figure \ref{FIG:nsvst} the monthly measured number of sunspots
is plotted versus time. This plot includes $44,444$ data points.
The approximate 11 years cycle can clearly be seen from this plot.

\section{Analysis of the Time Series}

The analysis of a given temporal signal with apparently random
fluctuations starts by asking if the value of the signal in a
given instant has any correlation with the signal in a later time.
The standard statistical tool for describing the signal is the
temporal correlation function and the corresponding power spectrum
or frequency spectrum also called spectral density. In general,
random time dependent perturbations give rise to random noise
characterized by a frequency spectrum following a power law
$1/f^{\beta}$ with $-2\leq \beta\leq 0$. Such a power law is found
in a rich variety of physical systems exhibiting some kind of
episodic activity. When $\beta \rightarrow 1$, the correlation
times grow strongly. This behavior is closely related to the
concept of fractal and self-similarity. In many cases, the
dynamics of the corresponding physical system is attracted towards
a critical state and the system is said to present self-organized
criticality.

The concepts of fractal and self-similarity are directly related
\cite{MAN}. Fractals are self-similar under an isotropic reduction or
scale transformation.

In many cases of interest, the concept of self-affine fractality
or self-affinity has shown to be very fruitful. In this "weak"
form of similarity, the sequence of data looks alike
when the reduction of the time scale is made with a different scaling
factor than
the one used for the measured variable. This means invariance under
anisotropic scale changes.
The random walk or
Brownian movement defined by a function $B(t)$ that determines the
instant position of the particle is a well-known example.
If $B(0)=0$, then at any time $t$
the characteristic function of the random walk verifies
$|\mu|^{-H_1}\,B(\mu\,t)$ for any $\mu \neq 0$. The Brownian exponent
is precisely $H_1=1/2$.
 In any case it is possible to
characterize the fractal behavior of a system through a
non-integer dimension usually called fractal dimension $D$,
related to the above mentioned $H_1$.

 An alternative characterization of fractal time series is by means of the
 roughness or Hurst exponent $H_u = 1 + H_1$ that measures the
persistency of the fluctuations related to the time series
\cite{PET}.

There have been different proposals for measuring these exponents.
One of them is the analysis of the variance of increments at a
given lag, also known with the geophysical name of semivariogram
\cite{BUR}. The detrended fluctuation analysis DFA \cite{DFA} has
shown to be an alternative procedure, some times a better one, for
analyzing fluctuations in time series and a way of defining
another similar exponent $H_a$. There are other possibilities as
the mobile average analysis introduced in reference \cite{VAN2}.

The statistical properties of the sunspot number fluctuations,
were analyzed by means of the different tools just mentioned, plus
the jump size distribution.

\subsection{Analysis of the Frequency Power Spectrum}

The analysis of correlation in time series starts with the
temporal correlation function defined by
\begin{equation}
G(\tau) = \langle x(t_0)\,x(t_0 + \tau)\rangle_{t_0} -
 \langle x(t_0)\rangle^2_{t_0}
\end{equation}
Clearly, whenever no temporal correlation is present, one has
$G(\tau) =0$. On the other hand, the rate with which $G(\tau)$
decreases from the average value of the instantaneous fluctuation
towards zero measures the correlation time that evaluates the
memory effects present in the signal.

The frequency spectrum is defined in terms of the amplitude of the
Fourier transform of the time signal squared, namely
\begin{equation}
S(f) = \lim_{T\rightarrow \infty}
\frac{1}{T}\,\left|\int^T_{-T}d\tau\,x(t)
\,e^{2\,i\,\pi\,f\,\tau}\right|^2
\end{equation}
In the particular case of a stationary process, i.e. its behavior
is independent of the particular instant chosen to start the
observation, the last expression ends in the cosine Fourier
transform
\begin{equation}
S(f) = 2\,\int_0^{\infty} d\tau\,G(\tau)\,\cos(2\pi f\tau)
\end{equation}
In general, a random time dependent fluctuation gives rise to a
random noise characterized by a frequency spectrum following a
power law
\begin{equation}
S(f) \propto \frac{1}{f^{\beta}}
\end{equation}
 over a large range of frequencies, with well known examples in the
 range $-2\leq \beta \leq 0$.

There exists a heuristic argument showing the particular nature
of the $\beta = 1$ fluctuations that defines the so called $1/f$
noise. If $S(f)\propto1/f^{\beta}$ and
$G(\tau)\propto1/\tau^{\gamma}$, then the cosine Fourier transform
above roughly shows that $1/f^{\beta}$ has to be proportional to
$1/f^{1-\gamma}$. Consequently, when $\beta$ approaches one,
$\gamma$ has to vanish. This means that for $\beta = 1$, the
assumed form of $G(\tau)$ is replaced by a logarithmic behavior
implying a much lower rate of decreasing of the correlation i.e.
correlation times grow strongly.

Another case of particular interest is the exponent $\beta = 5/3$
corresponding to the homogeneous turbulence in fluids. The analogy
is related to the presence of hierarchical processes like cascades
from large to small scale fluctuations.

\begin{figure}[htbp]
\vspace*{13pt}
\centerline{\psfig{file=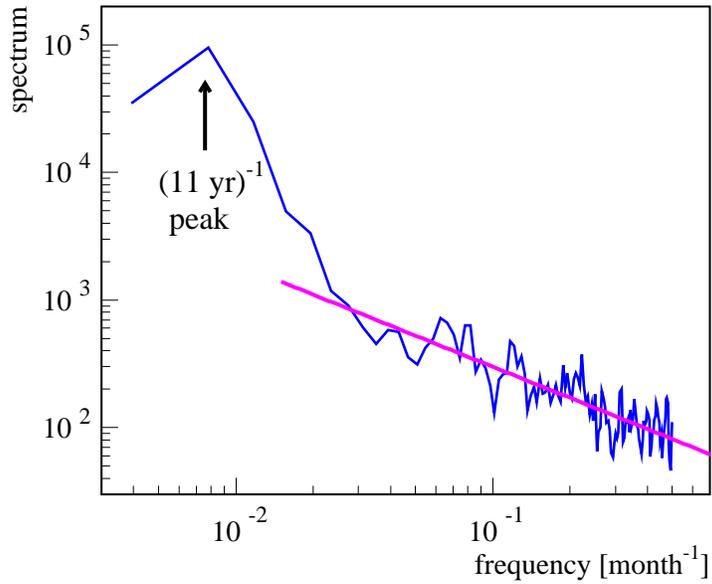}}
\caption{Power spectrum of the sunspot number fluctuation. The arrow
  indicates the peak at $f = (11\; {\rm yr})^{-1} = 0.0076\; {\rm
  month}^{-1}$.}
\label{FIG:spect}
\end{figure}

The analysis of sunspot data along the lines just described
provided an exponent clearly compatible with $\beta = 1$ as it is
shown in figure \ref{FIG:spect}.

In order to improve the analysis, a standard Butterworth filter
\cite{elec} ( $1/[1+\omega^{2\,n}]$ ) for $n=6$ and $8$, was
applied to the data in order to separate the evident low
 frequency components of the signal
 under consideration from the high frequency ones. The low
 frequency part shows clearly the
 well known $\sim 11$ year period of activity as a sharp peak in the
 corresponding spectrum, while
 the high frequency component has the noisy behavior presented in
figure \ref{FIG:spect}.  A power spectrum study of this high
frequency part provides the result

 \begin{equation}
 P(f)\approx f^{-0.8 \pm 0.2} \nonumber
 \end{equation}
 for
\[
 1/(132\, \mathrm{month}) < f < (1/1\, \mathrm{month}) \nonumber
\]
or alternatively \begin{equation}
  1.3\,\times\, 10^{-10}\,\mathrm{Hz} < f < 4\,\times\,10^{-7}\,\mathrm{Hz}  \nonumber
\end{equation}
We then conclude that the power spectrum of the sunspot number
temporal fluctuations, along more than two hundred years, is well
represented by a $1/f$ behavior.

\subsection{Detrended Fluctuation Analysis (DFA)}

 DFA \cite{DFA} is an alternative statistical method to classify temporal
correlations in a time series. It consists in dividing the time
series of $n$ values $x(t)$ into $n/s$ nonoverlapping groups, each
with $s$ values. Inside each group, the local trend or tendency
$y(t) = a\,t + b$ is defined as usual through the linear
least-square fit of the values in the group. From this, the
detrended fluctuation function $F(s)$ is defined as
\begin{equation}
\left[F(s)\right]^2=
\frac{1}{s}\,\sum^{(k+1)s}_{t=ks+1}\,\left[x(t) -
y(t)\right]^2\,\,\,\,;\,\,\,\,k=0,1,\ldots,\left(\frac{n}{s}
-1\right)
\end{equation}
Finally, the average of $F(s)$ over the $n/s$ intervals is
computed.

Whenever the time series data $x(t)$ is randomly uncorrelated (or
short range correlated), the expected behavior for the average is
a power law, namely
\begin{equation}
\langle F(s)\rangle \sim s^{H_a}
\end{equation}
with an exponent $H_a = 1/2$, characteristic value of a random
walk. If for a given range of $s$ the exponent $H_a$ happens to be
different from $1/2$, it detects the presence of long-range
correlations in that range. When $H_a
> 1/2$ there is a persistent behavior or a positive correlation
between the increments, while if $H_a < 1/2$ the behavior is
antipersistent showing the presence of a negative correlation. In
both cases it is customary to speak, after Mandelbrot, about
fractional Brownian movement.
\begin{figure}[htbp]
\begin{center}
\centerline{\psfig{file=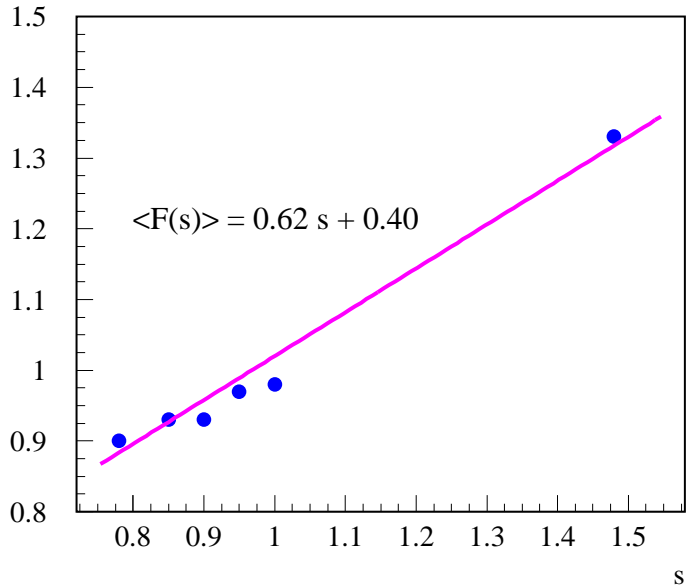}}
 \end{center}
\caption{Average of the detrended fluctuation.} \label{FIG:detre}
\end{figure}

The DFA avoids inherent trends at all scales and allows to easily
probe local correlations. Notice also that $H_a$ is related to the
fractal dimension through
\begin{equation}
D = 2 - H_a
\end{equation}
if the distribution is described in a two dimensional space as is
the case for the time series under analysis.

The behavior of the detrended fluctuation function for sunspot
data is shown in figure \ref{FIG:detre}. We found the presence of
a persistent behavior characterized by the exponent $H_a = 0.62$.

\subsection{Jump-Size Distribution Method }
\label{SEC:jump}

A further analysis of the time series can be performed in terms of
the jump-size distribution. The main objective is to determine the
probability distribution of the logarithm of consecutive relative
jumps in the number $S_i$ of sunspots, namely
\begin{equation} \label{eq:Ldef}
 L_{i} = \log\left|
\frac{S_{i}}{S_{i+1}} - 1\right| \end{equation} together with the
corresponding distribution \begin{equation} \label{eq:FLdef0}
F_{L} = Prob(\overline{L} \le L ) \end{equation}

\begin{figure}[htbp]
\vspace*{13pt}
\centerline{\psfig{file=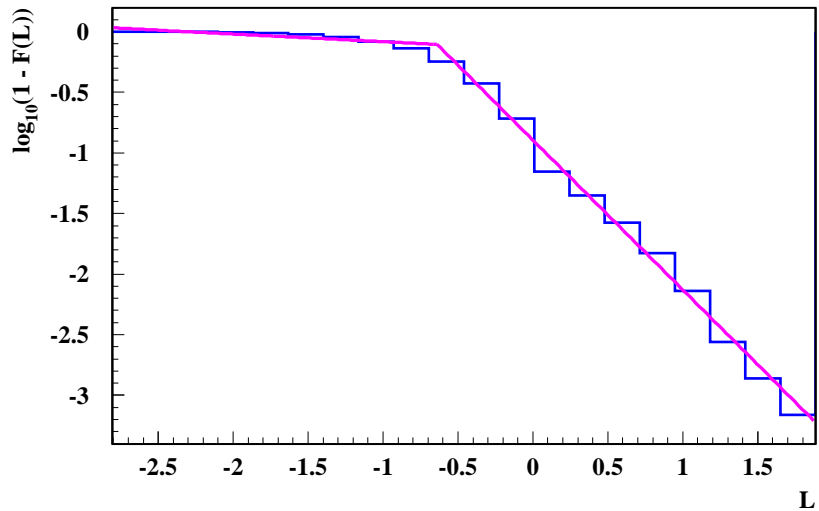}}
\caption{Logarithm of the cumulative distribution of consecutive
relative jumps, $\log_{10}[1-F_{L}(L)]$ (see equation
(\ref{eq:FLdef0})), plotted versus $L$ for the case of the number
of sunspots}
 \label{FIG:reldifdis}
\end{figure}

In figure \ref{FIG:reldifdis}, $\log_{10}[1 - F_{L}(L)]$ is
displayed for the case of the sunspot number using the data
plotted in figure \ref{FIG:nsvst}. It is found that $F_{L}$ shows
an inverse power behavior of the Zipf type \cite{zipf}, i.e., a
straight line on a double logarithmic plot, showing that
$[1-F_{L}(L)]$ can be expressed as the power
\begin{equation}
1-F_{L}(L) = L^{-\tau}
\end{equation}

We have found two different regimes: The one corresponding to negative
$L$ with $\tau_1 \approx 0$, and the one corresponding to positive $L$,
characterized by
\begin{equation}
 \tau_{2}  = 1.23 \pm 0.20 .
\end{equation}
 In the Appendix we have included an analysis of the
probability distribution of relative increments, just to
explicitly show that this power behavior should be expected
particularly in the case of a probability distribution of events
following a power law.

\section{Final Comments}

It is well-known that the $1/f$ noise is one of the most common
and ubiquitous features in Nature.  Superposition of effects
giving rise to signals with scale invariant distributions of
correlation times, the so-called scale similarity, could be on the
basis of the observed behavior. However, a proper explanation for
such a behavior is still lacking and for that reason, the physical
origin of the $1/f$ noise is pretty much an open question.
Nevertheless, a recent analysis \cite{rios} of models presenting self-organized
criticality seems to indicate that the origin of the $1/f$
behavior could be found in the superposition of local
power-spectra with characteristic frequencies suitably distributed
in space and having spectra that do not present the $1/f$
behavior. It has been also shown, in the mentioned context, that
one of the main features relevant for describing $1/f$ noise is
the well-known process of activation/deactivation.\footnote{For
  description of different examples of these {\em avalanche\/} type of
  processes, in connection with $1/f$ noise, see for instance
  \cite{rios,btw,weissman,millermiller}.}

 We find that both
mentioned ingredients are present in the case of sunspots. In
fact, the noise in the magnetic field measured at earth, for short
times (large frequencies) even during magnetic storms, presents a
$1/f^2$ spectrum. Consequently, the property of the sunspots
spectrum can be related to the interference of random distributed
magnetic disturbances. Moreover, the commonly accepted mechanism for
the sunspot appearance is an activation/deactivation
process, as it was discussed in the introduction.

With reference to the roughness exponent, one can conclude that it
also detects the persistent behavior or positive correlation
between increments, due to the physical mechanisms giving rise to
sunspots.

\section*{Acknowledgments}
H. F., C. A. G. C. and S. J. S. are partially supported by the
Consejo Nacional de Investigaciones Cient\'{\i}ficas y T\'ecnicas
(CONICET), and Agencia Nacional de Promoci\'on Cient\'{\i}fica y
Tecnol\'ogica, Argentina. C. H. thanks Fulbright Commission and
Fundaci\'{o}n Antorchas for financial help. C. A. G. C. thanks J.
Bernabeu and the Departamento de F\'{\i}sica Te\'orica, Universidad de
Valencia for the warm hospitality extended to him.

\appendix
\renewcommand{\theequation}{A.\arabic{equation}}
\setcounter{equation}{0}
\section*{Appendix: On Probability Distribution of Relative Increments }

In this Appendix we illustrate how  the uniform and the power
distributions  give rise to the probability distribution of
relative increments of the Zipf-like type. To start with, let us
consider a random variable $X$ with probability distribution
$f_X(x)$ ($-\infty < x <\infty$). The distribution $f_X$ is
normalized, that is,
\begin{equation}
\int_{-\infty}^{\infty} f_X(x) dx = 1.
\end{equation}
The cumulative probability distribution $F_X(x)$ is defined via:
\begin{equation}
F_X(x) = \int_{-\infty}^{x} f_X(z) dz.
\end{equation}
$F_X(x)$ gives the probability that $X\le x$, and clearly,
$f_X(x)= dF_X(x)/dx$. Let $x_i$, $i=1,2,\ldots$, be a series of
independent instances of $X$, that is, each $x_i$ is an
independent sample of the distribution $f_X(x)$. We are interested
in calculating the probability distribution of certain functions
of these numbers, namely, $r_i$, $R_i$ and $L_i$, $i=1,2,\ldots$
defined as
\begin{eqnarray}
\displaystyle r_i &=&\displaystyle {{x_i}\over{x_{i+1}}} - 1
\\*[4pt]
R_i &=& |r_i|\\ L_i &=& \ln R_i
\end{eqnarray}

The probability distribution of the $r_i$'s can be evaluated as
follows (overlined symbols refer to random variables):
\begin{eqnarray}
F_r(r) &=& \hbox{Prob}(\overline{r}\le r)\\
       &=&\displaystyle\hbox{Prob}\!\left({{x_i}\over{x_{i+1}}} - 1\le
 r\right)\\*[6pt]
       &=&\displaystyle\hbox{Prob}(x_i\le(r+1) x_{i+1}).
\end{eqnarray}
Since $x_i$ and $x_{i+1}$ are independent, the probability of the
last equation can be evaluated straightforwardly
\begin{equation}
F_r(r) = \hbox{Prob}(x_i\le(r+1) x_{i+1}) =
\int_{-\infty}^{\infty} f_X(x_{i+1})
F_X\left((r+1)x_{i+1}\right)\, dx_{i+1}.
\end{equation}
Or, equivalently,
\begin{equation}                                \label{eq:Frdef}
F_r(r) = \int_{-\infty}^{\infty}dx_1 f_X(x_1)
         \int_{-\infty}^{(r+1)x_1} dx_2 f_X(x_2).
\end{equation}
The probability density distribution, $f_r$, can be then evaluated
via:
\begin{equation}                                 \label{eq:frdef}
f_r(r) = {{dF_r(r)}\over{dr}} =
\int_{-\infty}^{\infty} dx\> x\,f_X(x)\,f_X((r+1)x) .
\end{equation}
The distributions for $R$ and $L$ can be calculated similarly:
\begin{equation}
F_R(R) = \hbox{Prob}(\overline{R}\le R) = \hbox{Prob}(|r|\le R).
\end{equation}
It is evident that $F_R(R) = 0$ for $R\le 0$. For $R>0$ one has,
\begin{equation}                                \label{eq:FRdef}
F_R(R) = \hbox{Prob}(-R\le r \le R) = F_r(R) - F_r(-R).
\end{equation}
On the other hand,
\begin{eqnarray}
F_L(L) &=& \hbox{Prob}(\overline{L}\le L)\\
       &=& \hbox{Prob}(\ln R\le L)\\
       &=& \hbox{Prob}(R \le e^{L}),
\end{eqnarray}
therefore,
\begin{equation}                                \label{eq:FLdef}
F_L(L) = F_R(e^{L}).
\end{equation}
The probability density distribution can be then calculated
similarly as in equation (\ref{eq:frdef}).

Let us now evaluate these probability distributions for the
particular cases of the uniform and the power distributions.

\subsubsection*{The uniform distribution}

The uniform distribution is given by
\begin{equation}
f_X(x) = \left\{ \begin{array}{ll} 1 &\hbox{if\ } 0\le x \le 1
\\*[16pt]
0 & \hbox{otherwise}
\end{array}\right.
\end{equation}

Applying equation (\ref{eq:Frdef}), one obtains
\begin{equation}
F_r(r) = \left\{\begin{array}{ll} 0 & \hbox{if\ } r<-1 \\*[16pt]
\displaystyle {{r+1}\over{2}} & \hbox{if\ } -1 \le r \le
0\\*[16pt] \displaystyle {{2r+1}\over{2(r+1)}} & \hbox{if\ } r > 0
\end{array}\right.
\end{equation}

And then from equation (\ref{eq:FRdef})
\begin{equation}        \label{eq:uniFR}
F_R(R) = \left\{\begin{array}{ll} 0 & \hbox{if\ } R\le 0 \\*[16pt]
\displaystyle {{2R+1}\over{2(R+1)}}-{{1-R}\over{2}} &
 \hbox{if\ } 0 < R < 1\\*[16pt]
\displaystyle {{2R+1}\over{2(R+1)}} & \hbox{if\ } R \ge 1
\end{array}\right.
\end{equation}

\subsubsection*{The power distribution}

\noindent Consider the distribution
\begin{equation}
f_X(x) = \left\{ \begin{array}{ll} (\alpha-1)x^{-\alpha}
&\hbox{if\ } x>1 \\*[16pt] 0 & \hbox{otherwise}
\end{array}\right.
\end{equation}
where $\alpha$ is a real parameter ($\alpha>1$).

Applying equation (\ref{eq:Frdef}), one obtains
\begin{equation}
F_r(r) = \left\{\begin{array}{ll} 0 & \hbox{if\ } r<-1 \\*[16pt]
\displaystyle (r+1)^{\alpha}\left({1\over{r+1}}-{1\over2}\right) &
 \hbox{if\ } -1 < r < 0\\*[16pt]
\displaystyle 1- {1\over{2(r+1)^{\alpha-1}}} & \hbox{if\ } r > 0
\end{array}\right.
\end{equation}

And then from equation (\ref{eq:FRdef})
\begin{equation}   \label{eq:powFR}
F_R(R) = \left\{\begin{array}{ll} 0 & \hbox{if\ } R\le 0 \\*[16pt]
\displaystyle 1-{1\over{2(R+1)^{\alpha-1}}}-
(1-R)^{\alpha}\left({1\over{1-R}}-{1\over{2}}\right) &
 \hbox{if\ } 0 < R < 1\\*[16pt]
\displaystyle 1-{1\over{2(R+1)^{\alpha-1}}} & \hbox{if\ } R \ge 1
\end{array}\right.
\end{equation}

\begin{figure}[htbp]
\vspace*{13pt}
\centerline{\psfig{file=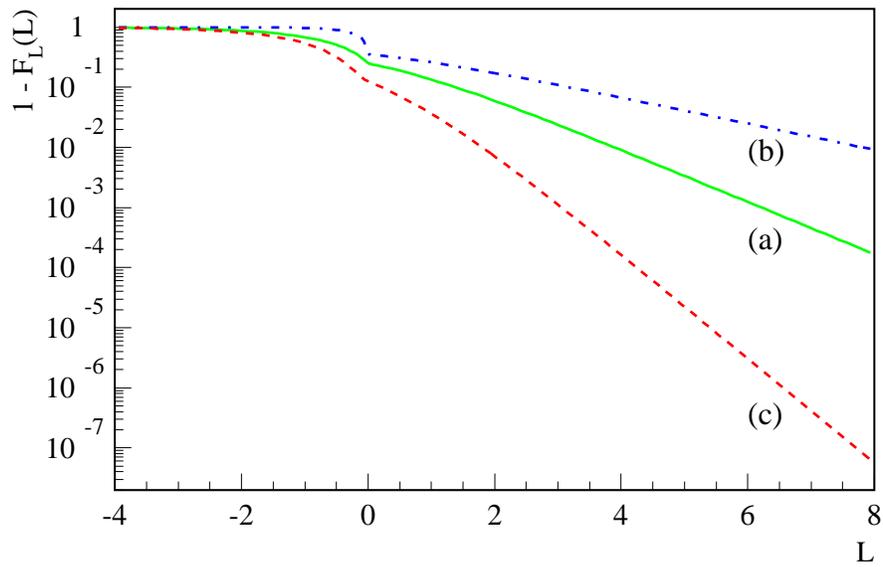}}
\caption{Some cumulative functions ($1 -F_L(L)$) plotted versus $L$.
 (a) The uniform distribution (equation (\ref{eq:uniFR})). (b) The
 power distribution (equation (\ref{eq:powFR})) in the case $\alpha =
 3/2$. (c) Same as (b) but for $\alpha = 3$. In every case two
 asymptotic regimes are clearly seen.}
\label{FIG:acupowuni}
\end{figure}

In these two examples, two different straight line regimes are clearly
present, as it is illustrated in figure \ref{FIG:acupowuni} where
$1-F_L(L)$ (see equation (\ref{eq:FLdef})) is plotted as a function of $L$.
 This is precisely the
behaviour of the corresponding distribution in the case of the sunspot
number, as shown in section \ref{SEC:jump} (figure \ref{FIG:reldifdis}).


\clearpage


\begin{thebibliography}{99}

\bibitem{bray}
see for example {\em Sunspots\/} by R.J. Bray, R.E. Loughhead,  Dover
Publications, New York 1979.
\bibitem{matt}
W.H. Matthaeus, M.L. Goldstein, {\em Phys. Rev. Lett.,\/} {\bf 57}, 495
(1986).
\bibitem{data}
http://www.oma.be/KSB-ORB/SIDC/index.html
\bibitem{MAN} B.B. Mandelbrot, The Fractal Geometry of Nature,
Freeman, N.York (1982).
\bibitem{PET} E.A. Peters, Chaos and Order in the Capital Markets,
J. Wiley (1991).
\bibitem{BUR} P.A. Burrough, {\em Nature,\/} {\bf 294}, 240 (1981).
\bibitem{elec} A. Zverev, Handbook of Filter Synthesis, John Wiley
  (1967). 
\bibitem{DFA} C.K. Peng {\em et al.,} {\em Phys. Rev. E,\/} {\bf 49},
  1685 (1994).
\bibitem{VAN2} N. Vandewalle, M. Ausloos, {\em Phys. Rev. E,\/} {\bf 58},
6832 (1999).
\bibitem{zipf}
see for example ``How Nature Works" by P. Bak, Oxford University
Press, 1997.
\bibitem{rios}
P. De Los Rios, Yi-Cheng Zhang, {\em Phys. Rev. Lett.,\/} {\bf 82},
472 (1999).
\bibitem{btw} P. Bak, C. Tang, K. Wiesenfeld, {\em
  Phys. Rev. Lett.,\/} {\bf 59}, 381 (1987).
\bibitem{weissman} M. B. Weissman, {\em
  Rev. Mod. Phys.,\/} {\bf 60}, 537 (1988).
\bibitem{millermiller} S. L. Miller, W. M. Miller, P. J. McWhorter, {\em
  J. Appl. Phys.,\/} {\bf 73}, 2617 (1993).
\end{thebibliography}
\end{document}